\documentclass[twoside,twocolumn,english,aps,prl,superscriptaddress,prx]{revtex4-1}
\usepackage[latin9]{inputenc}
\usepackage{amsmath}
\usepackage{amssymb}
\usepackage{graphicx}
\usepackage{esint}
\makeatletter
\usepackage{dcolumn}
\usepackage{bm}
\usepackage{xcolor}
\usepackage{babel}
\usepackage[colorlinks, citecolor=blue ]{hyperref}

\makeatother

\usepackage{babel}
\begin{document}

\title{Twofold quadruple Weyl nodes in chiral cubic crystals}


\author{Tiantian Zhang}
\affiliation{Department of Physics, Tokyo Institute of Technology, Ookayama, Meguro-ku, Tokyo 152-8551, Japan}
\affiliation{Tokodai Institute for Element Strategy, Tokyo Institute of Technology, Nagatsuta, Midori-ku, Yokohama, Kanagawa 226-8503, Japan}

\author{Ryo Takahashi}
\affiliation{Department of Physics, Tokyo Institute of Technology, Ookayama, Meguro-ku, Tokyo 152-8551, Japan}

\author{Chen Fang}
\affiliation{Institute of physics, Chinese academy of sciences}
\affiliation{CAS Center for Excellence in Topological Quantum Computation, Beijing 100190, China}

\author{Shuichi Murakami}
\affiliation{Department of Physics, Tokyo Institute of Technology, Ookayama, Meguro-ku, Tokyo 152-8551, Japan}
\affiliation{Tokodai Institute for Element Strategy, Tokyo Institute of Technology, Nagatsuta, Midori-ku, Yokohama, Kanagawa 226-8503, Japan}

\begin{abstract}
{Conventional Weyl nodes are twofold band crossings that carry a unit monopole charge, which can exist in condensed matter systems with the protection of translation symmetry. 
Unconventional Weyl nodes are twofold/multifold band crossings carrying a quantized monopole charge larger than one, and their existence needs the protection of additional crystalline symmetries.
Studies on unconventional Weyl nodes are already very comprehensive, such as twofold Weyl nodes with Chern number of $C=2/C=3$ and fourfold/sixfold Weyl nodes with $C=4$. 
Yet in this paper, we propose a newfound twofold unconventional Weyl node with $C=4$, which can exist in any  chiral cubic systems with integer spin. 
Such kind of twofold quadruple Weyl node has a cubic band dispersion along [111] direction and will  evolve into a fourfold quadruple Weyl node after considering spin-orbit coupling. 
In this paper, we exhaust all the possible chiral cubic space groups and corresponding $k$-points, which can have twofold quadruple Weyl nodes. We also propose a series of LaIrSi-type materials that both have twofold quadruple Weyl nodes in electronic systems and the phonon spectra.}
\end{abstract}

\maketitle

\section{Introduction}
Unconventional chiral fermions\cite{DWP_NEWFERMIONS,DWP_DOUBLE_HUANG,DWP_MULTI_INPHOTONIC,double_benjamin,PhysRevX.6.031003,double_bouhon} have been well studied owing to the lack of the high-energy counterparts and hosting intriguing physical properties, such as negative magnetoresistance due to ``chiral anomaly''\cite{chiral1,chiral2,chiral3,chiral4,cano2017chiral,jia2019observation}, chiral zero sound\cite{PhysRevX.9.021053,PhysRevX.9.031036}, photo induced anomalous Hall effect\cite{PhysRevLett.116.026805}, quantized circular photogalvanic effect\cite{QCPE_NC,PhysRevB.96.075123,rees2019observation,CPGE_PRB,ji2019spatially}, topologically protected Fermi-arc surface states\cite{Xu613,Xu2015,Deng2016,lv2015experimental,Lu622,arcs_wte2}, and so on. 
Because of the interesting physical properties, studies for unconventional Weyl nodes also have been widely studied in bosonic systems in the last decade, such as phonons\cite{zhang2018double,miao2018observation,stenull2016topological}, photons\cite{Lu622,DWP_MULTI_INPHOTONIC}, magnons\cite{chisnell2015topological,yao2018topological,li2017dirac}, and so on. 
{Aside from the unconventional Weyl nodes, an interesting one called ``Kramers Weyl node'', which can only exist in chiral crystals with time-reversal symmetry, also receives wide attention for its unique feature. Kramers Weyl nodes are band crossings located at time-reversal-invariant momenta (TRIM) and the chirality for the same Kramers Weyl node depends on the chirality of crystal structure\cite{chang2018topological,zhang2017ultraquantum,wan2018theory}. Since Kramers Weyl nodes are 
irreducible representations at TRIM, they will be more robust to the perturbations than Weyl nodes locating at non-TRIM and not easy to be created/annihilated in pairs by modulating the parameters of a system. Weyl nodes locating at non-TRIM are usually related to each other by symmetries\cite{weng2015weyl,yang2015weyl,lv2015experimental,huang2015weyl,sun2015prediction,soluyanov2015type,wu2016observation,wang2016observation}, yet Kramers Weyl nodes can appear alone without relating to the Weyl nodes with opposite chirality.} 
Thus, a pair of non-Kramers Weyl nodes with opposite chirality will have an equal energy, while Kramers Weyl nodes are free to have different energies. 
Such kind of energy offset makes Kramers Weyl nodes to be ideal platforms for observing quantized circular photo-galvanic effect (CPGE) in electronic systems, where the induced photo current is proportional to the monopole charge of the Weyl nodes\cite{QCPE_NC,PhysRevB.96.075123,rees2019observation,CPGE_PRB}. 


\begin{center}
\begin{figure}
\includegraphics[scale=1.3]{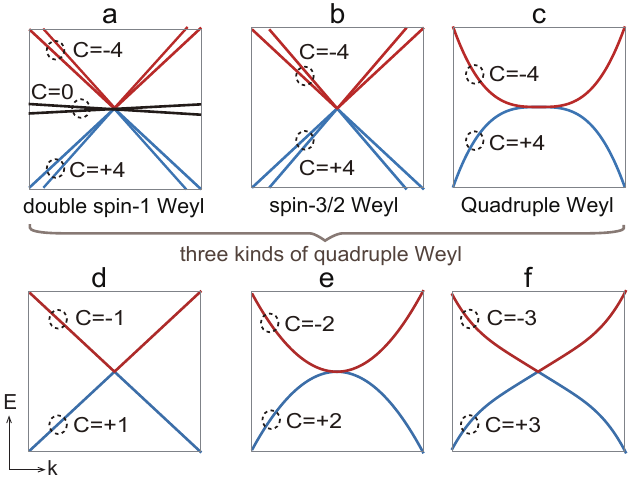}
\caption{Three kinds of Weyl nodes with a monopole charge of $+4$ (a-c) and four kinds of twofold Weyl node with a different Chern number (c-f). (a) Sixfold degenerated double spin-1 Weyl node, which is a composition of two identical spin-1 Weyl nodes. (b) Fourfold degenerated spin-$\frac{3}{2}$ Weyl node. The Chern number for each band is $+1, +3, -1, -3$. (c) Newly discovered quadruple Weyl node with a twofold degeneracy. The Chern number for each band is $+4, -4$. (d-f) are other three kinds of twofold Weyl nodes that can exist in crystalline materials, which are called single Weyl node, double Weyl node, and triple Weyl node, separately. 
\label{fig:FIG_0}}
\end{figure}
\end{center}

Although researches on unconventional Weyl nodes, especially the Kramers Weyl nodes, are already very comprehensive, they do not exhaust all the unconventional Weyl nodes, formed by two bands. {In this paper, we expand the definition of Kramers Weyl node to both spinful and spinless systems with time-reversal symmetry ($\mathcal{T}$).} 
A twofold Kramers Weyl node with Chern number of $\pm4$ is proposed in this paper for the first time. Such kind of twofold quadruple Weyl node can exist in all chiral cubic integer-spin systems with $\mathcal{T}$, and is located at TRIM (We will explain the reasons why it has been missed in the following text). We also propose a series of easily synthesized material candidates that have the twofold quadruple Weyl node in the electronic structure and those in the phonon spectra based on first-principle calculations. All the material candidates have a similar structure with LaIrSi, and some of them are well-known superconductors\cite{CHEVALIER1982801,JPSJ2018}.

\section{Quadruple Weyl nodes}
\subsection{Quadruple Weyl nodes in condensed matter physics}
The existence of unconventional Weyl nodes in condensed matter physics benefits from a lower symmetry than the Poincar\'e groups in high-energy physics, which only preserves spin-$\frac{1}{2}$ Weyl nodes\cite{weng2015weyl,chiral1,chiral2,chiral3,chiral4,SWP_LR}. 
Among space-group-allowed unconventional Weyl nodes, the ones with a large Chern number received the most attention. For example, unconventional chiral fermions with Chern number of $\pm4$, $i.e.$, quadruple Weyl fermion, have been widely studied in transition metal silicides\cite{chang2017unconventional,PhysRevLett.119.206402}, as shown in Fig.~\ref{fig:FIG_0} (a-b).
 Figure~\ref{fig:FIG_0} (a) is a sixfold degenerated quadruple Weyl fermion consisting of two identical spin-1 Weyl nodes, so it is also called double spin-1 Weyl fermion. Figure~\ref{fig:FIG_0} (b) is a fourfold Rarita-Schwinger-Weyl~ (RSW) fermion\cite{RSW}, having a Chern number of $+1$, $+3$, $-1$, $-3$ for each band. The RSW fermion is also named as spin-$\frac{3}{2}$ Weyl fermion, because its low-energy effective Hamiltonian can be written as $H=\mathbf{k}\cdot\mathbf{S}$, where $\mathbf{S}$ is the rotation generators for spin-$\frac{3}{2}$ particles. 
Figure~\ref{fig:FIG_0} (c) is the new quadruple Weyl fermion with twofold degeneracy, having a large Chern number of $\pm 4$ for each band. 
Although the unconventional Weyl fermions with twofold degeneracies have been studied, previous studies have been limited to double Weyl nodes with Chern number of $\pm 2$ and triple Weyl nodes with Chern number of $\pm 3$, as shown in Fig.~\ref{fig:FIG_0} (e-f), and studies on twofold degenerated quadruple Weyl nodes were missed. 
To the best of our knowledge, this is the first time for people to realize that a quadruple Weyl node can be obtained in twofold degenerated bands, which is also the highest Chern number that a twofold degenerated Weyl node can have in crystalline materials. 
Here we would like to note that twofold quadruple Weyl node can only be obtained in materials with an integer spin and chiral cubic symmetry, $i.e.$, the point groups $T$ and $O$ with $\mathcal{T}$, such as spinless electron, phonon, photon, magnon, and so on.

\subsection{Why missed it}
\label{IIB}
Although a  tremendous effort has been made in the studies of unconventional Weyl nodes, most of them focus on multifold ones, such as threefold spin-1 Weyl nodes\cite{rao2019observation,takane2019observation,miao2018observation}, fourfold spin-$\frac{3}{2}$ points\cite{PhysRevLett.119.206402,chang2017unconventional,furusaki2017weyl}, sixfold double spin-1 Weyl nodes\cite{PhysRevLett.119.206402,chang2017unconventional,furusaki2017weyl}, eightfold double Dirac nodes\cite{DWP_NEWFERMIONS,double_benjamin}, and so on\cite{wu2019higherorder}. 
Systematic studies on twofold unconventional Weyl nodes have also been implemented. However, it stopped at double Weyl nodes\cite{DWP_fang,PhysRevB.96.045102,T_Nagaosa_2014,jian2015double} and triple Weyl nodes\cite{DWP_fang,PhysRevB.96.045102,T_Nagaosa_2014} in the presence/absence of different rotation symmetries, $\mathcal{T}$, and SU(2) symmetry. The twofold quadruple Weyl node was overlooked in previous works since the cubic symmetry is not taken into consideration, so the largest Chern number of a 2-band system is $\pm 3$ in the previous studies.

In general, for a twofold Weyl node with Chern number C=$N$, an example of the low-energy effective $k\cdot p$ model without considering additional crystalline symmetries can be written as:
$H (\mathbf{k})=\begin{pmatrix}
Ak_{z} & B (k_{x}-ik_{y})^{N} \\ 
B (k_{x}+ik_{y})^{N}  & -Ak_{z} 
\end{pmatrix}$, where $A$ and $B$ are real constants. Thus, we can obtain a Weyl node with arbitrary Chern number if there is no restriction of crystalline symmetries. In the previous studies, only an influence of the rotation symmetries on the $k\cdot p$ model was considered, but not the cubic symmetry. Thus, a twofold double Weyl node with Chern number of $\pm 2$ and a twofold triple Weyl node with Chen number of $\pm 3$ were proposed in a system with $C_{3}$/$C_{4}$/$C_{6}$ rotation symmetry and in systems with $C_{3}$/$C_{6}$ rotation symmetry, respectively, as shown in Fig.~\ref{fig:FIG_0} (e-f). Under the $C_{3}$ rotation symmetry, if we further take cubic symmetry into consideration, a twofold quadruple Weyl node with Chern number of $\pm 4$ is possible, as shown in Fig.~\ref{fig:FIG_0} (c), which will be discussed below. 
{Since the twofold quadruple Weyl node needs the protection of chiral cubic symmetry and $\mathcal{T}$, their locations are restricted at $\Gamma$ point for all chiral cubic space groups and $ (\frac{\pi}{a},\frac{\pi}{a},\frac{\pi}{a})$ point for only seven chiral cubic space groups, where $a$ is the lattice constant for the primitive cell. Possible space groups and corresponding irreducible representations required for the twofold quadruple Weyl are listed in Tab~\ref{tab:QW}.}

In the following, we will discuss the twofold quadruple Weyl node in electronic systems and phononic systems separately, based on first-principle calculations.

\renewcommand\arraystretch{1.8}
\begin{table}[]
\begin{tabular}{@{}c|c|c|c|cl@{}}
\hline
k-point     & \multicolumn{2}{c|}{ (0,0,0)} & \multicolumn{2}{c}{$ (\frac{\pi}{a},\frac{\pi}{a},\frac{\pi}{a})$} \\ \hline
\hline
\begin{tabular}[c]{@{}c@{}}possible\\space groups  \end{tabular} &     \#195-\#199      & \#207-\#214          & \begin{tabular}[c]{@{}c@{}}\#195,\#197\\ \#199\end{tabular}            &  \begin{tabular}[c]{@{}c@{}}\#207,\#208\\  \#211,\#214  \end{tabular}         \\ \hline
NSOC        & $^{1}$E$^{2}$E            & E            & $^{1}$E$^{2}$E            & E           \\ \hline
SOC         & $^{1}$F$_{\frac{3}{2}}$$^{2}$F$_{\frac{3}{2}}$            & F$_{\frac{3}{2}}$            & $^{1}$F$_{\frac{3}{2}}$$^{2}$F$_{\frac{3}{2}}$            &F$_{\frac{3}{2}}$         \\ \hline
\end{tabular}
\caption{Irreducible representations for twofold quadruple Weyl nodes in spinless systems, and the corresponding irreducible representation after considering SOC. $a$ is the lattice constant for the primitive cell.}
\label{tab:QW}
\end{table}


\section{Quadruple Weyl fermions}
\subsection{Crystal structure and quadruple Weyl fermions in LaIrSi-type materials}
\begin{center}
\begin{figure}
\includegraphics[scale=0.82]{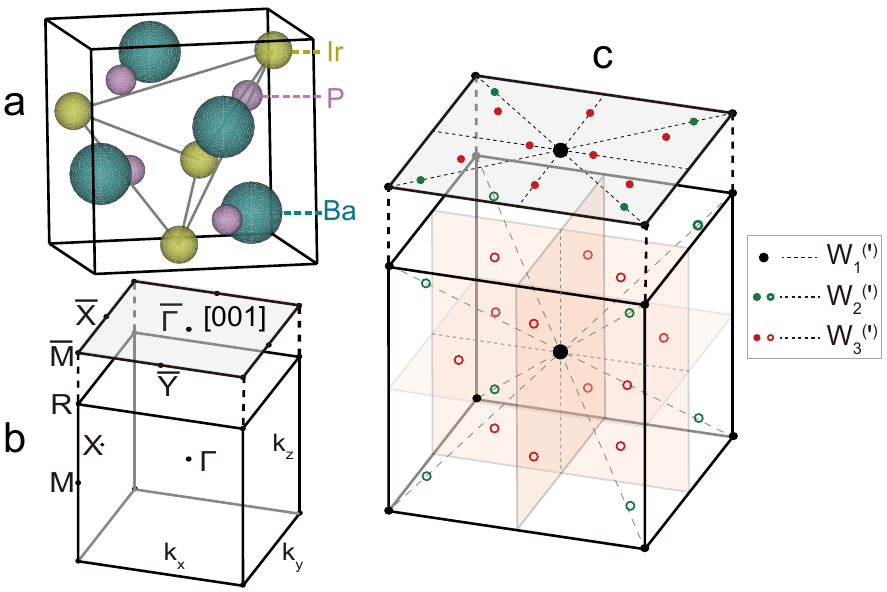}
\caption{Crystal structure and Weyl nodes for BaIrP. (a) Crystal structure for BaIrP, which is one of the LaIrSi-type materials. (b) Brillouin zone and surface Brillouin zone for BaIrP along [001] direction. (c) Positions of Weyl nodes for BaIrP both in the absence and in the presence of SOC. Positions for the Weyl nodes are accidentally quite close between the cases with and without SOC. Green circles are Weyl nodes $W_{1}$ ($W_{1}'$) along the $C_{3}$ preserved diagonal lines with a negative monopole charge in the spinless (spinful) case. Red circles/dots on the $k_{x,y,z}=0$ plane are Weyl nodes $W_{2}$ ($W_{2}'$) with a positive monopole charge without (with) SOC. The black dot represents the quadruple Weyl $W_{1}$ ($W_{1}'$) without (with) SOC.
\label{fig:FIG_1}}
\end{figure}
\end{center}

LaIrSi-type materials\cite{CHEVALIER1982801} contain 3 different kinds of elements and 12 atoms in one primitive cell, as shown in Fig.~\ref{fig:FIG_1} (a), preserving space group P2$_{1}$3~ (\#198). {Generators for LaIrSi-type materials are $\{C_{2}^{001}|\frac{1}{2}0\frac{1}{2}\}$, $\{C_{2}^{010}|0\frac{1}{2}\frac{1}{2}\}$ and $C_{3}^{111}$.} 
Brillouin zone~ (BZ) and the surface BZ along [001] direction for LaIrSi-type materials are shown in Fig.~\ref{fig:FIG_1} (b). In our work, we explore LaIrSi-type materials including LaIrSi, LaRhSi, CaPtSi, $X$PdSi, $X$IrP, $X$PtSi, $X$PtGe~ ($X$=Sr, Ba), which have quadruple Weyl nodes in both the electronic structure and the phonon spectra. In this section, we will focus on the quadruple Weyl fermions in LaIrSi-type materials and the evolution before and after considering spin-orbit coupling~ (SOC). All LaIrSi-type materials have a twofold degenerated quadruple Weyl node near the Fermi energy in the absence of SOC and a 4-fold degenerated quadruple Weyl after considering SOC in the electronic structure. Quadruple Weyl fermions in LaIrSi-type materials are contributed by the $e_{g}$ orbits from the lanthanide or transition-metal element. Near the Fermi energy, electronic bands are mainly contributed by unconventional Weyl fermions in both the absence and presence of SOC, so physical properties related to unconventional excitations are very likely to be observed. After calculations based on density functional theory (DFT)\cite{CAL_VASP}, we found that topological properties in all the LaIrSi-type materials are the same. Therefore in the following, we will pick up one material, $i.e.$, BaIrP, for further discussion. Electronic structure for other LaIrSi-type materials are listed in the supplementary.


\subsection{Chiral fermions in BaIrP without SOC}
\begin{center}
\begin{figure}
\includegraphics[scale=1]{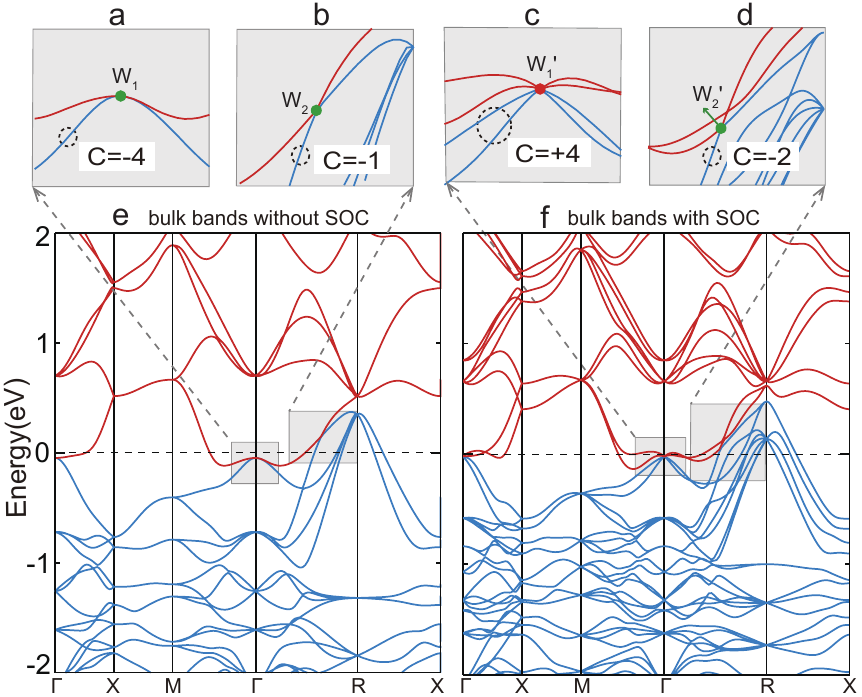}
\caption{Band structure of BaIrP with and without SOC. (a) is the quadruple Weyl node and (b) is the spin-$\frac{1}{2}$ Weyl node in BaIrP without SOC. (c) is the  spin-$\frac{3}{2}$ Weyl node and (d) is the twofold double Weyl node with a Chern number of $-2$ in BaIrP with SOC. (e-f) Bulk bands for BaIrP without and with SOC.
\label{fig:FIG_2}}
\end{figure}
\end{center}

In the absence of SOC, the twofold degeneracy node in BaIrP at $\Gamma$ point is a quadruple Weyl node $W_{1}$ with a monopole charge of $-4$ for the valence band. As shown in Fig.~\ref{fig:FIG_2}~ (a), the blue line is the valence band and the red line is the conduction band. Another degenerated node between the valence band and conduction band along $\Gamma$-R direction is a spin-$\frac{1}{2}$ Weyl node $W_{2}$ with a monopole charge of $-1$, as shown in Fig.~\ref{fig:FIG_2}~ (b). 
Here, we would like to note that there are 8 $W_{2}$ Weyl nodes in the BZ due to the symmetries of BaIrP, and their positions are listed in Tab.~\ref{tab:table} and marked by green circles in Fig.~\ref{fig:FIG_1} (c). The sum of the monopole charge for Weyl nodes $W_{1}$ and $W_{2}$ are $-12$, which means that there should be several Weyl nodes with a total monopole charge of $+12$ according to the no-go theorem\cite{NoGo}. 
After a numerical calculation, we find that there are 12 spin-$\frac{1}{2}$ Weyl nodes $W_{3}$, each of them carries a monopole charge of $+1$. Those 12 spin-$\frac{1}{2}$ Weyl nodes are distributed on the $k_{x,y,z}$=0 planes as listed in Tab.~\ref{tab:table} and marked by the red circles in Fig.~\ref{fig:FIG_1} (c). 
Figure~\ref{fig:FIG_1} (c) also shows the approximate positions for Weyl nodes on the surface BZ along [001] direction, where the green dots represent the Weyl nodes with a negative monopole charge and the red dots represent the Weyl nodes with a positive monopole charge. In the spinless case, the black dot at $\Gamma$/$\bar{\Gamma}$ is the quadruple Weyl node with a monopole charge of $-4$.

\begin{table*}[]
\begin{tabular}{cccc}
\hline 
Positions ($\frac{2\pi}{a}$) & $W_{1}$ ($W_{1}'$) & $W_{2}$ ($W_{2}'$)        & $W_{3}$ ($W_{3}'$)    \\
\hline \hline %
Monopole charge     &  $-4$ ($+4$)  & $-1$ ($-2$)      &  $+1$ ($+1$) \\
\hline  
NSOC                & (0,0,0)      & \begin{tabular}[c]{@{}c@{}}$\pm$ (0.318,0.318,0.318) $\pm$ (-0.318,-0.318,0.318) \\ $\pm$ (-0.318,0.318,-0.318) $\pm$ (0.318,-0.318,-0.318) \end{tabular} & \begin{tabular}[c]{@{}c@{}}$\pm$ (0.179,0.244,0) $\pm$ (0,0.179,0.244) $\pm$ (0.244,0,0.179) \\  $\pm$ (0,-0.179,0.244) $\pm$ (0.179,-0.244,0)  $\pm$ (-0.244,0,0.179) \end{tabular} \\ \hline 
SOC                 & (0,0,0)      & \begin{tabular}[c]{@{}c@{}}$\pm$ (0.277,0.277,0.277)  $\pm$ (-0.277,-0.277,0.277) \\$\pm$ (-0.277,0.277,-0.277) $\pm$ (0.277,-0.277,-0.277)  \end{tabular}& \begin{tabular}[c]{@{}c@{}}$\pm$ (0.174,0.3,0) $\pm$ (0,0.174,0.3) $\pm$ (0.3,0,0.174)\\$\pm$ (0,-0.174,0.3) $\pm$ (-0.3,0,0.174) $\pm$ (0.174,-0.3,0) 
\end{tabular} \\ \hline 
\end{tabular}
\caption{Positions for the Weyl nodes $W_{1}$ ($W_{1}'$), $W_{2}$ ($W_{2}'$), and $W_{3}$ ($W_{3}'$) in BaIrP without (with) considering spin-orbit coupling, where $a$ is the lattice constant.}
\label{tab:table}
\end{table*}

\subsection{Chiral fermions in BaIrP with SOC}
After considering SOC, the spin-$\frac{1}{2}$ Weyl node $W_{2}$ will evolve into a quadratic double Weyl node $W_{2}'$, which is a twofold double Weyl node with a Chern number of $\pm 2$ for each band, due to the presence of $C_{3}$ rotation symmetry along $\Gamma$-R direction and $\mathcal{T}$. Therefore the sum monopole charge contributed by the Weyl node $W_{2}'$ will  be $-16$, instead of $-8$ in the spinless case. However, every spin-$\frac{1}{2}$ Weyl node $W_{3}$ will split into two spin-$\frac{1}{2}$ Weyl nodes ($W_{3}'$ and $W_{3}''$) due to lack of rotation symmetry at the $W_{3}$ point, and they will contribute a total monopole charge of $+24$. Among those 24 Weyl nodes, 12 of them ($W_{3}'$) will be located around their original positions, and the other 12 Weyl nodes $W_{3}''$ will move close to $\Gamma$ point when SOC is gradually turned on, and will be annihilated with 12 Weyl nodes $W_{1}''$ coming from $\Gamma$ point. 
What happens to the quadruple Weyl node $W_{1}$ is the most interesting one. The quadruple Weyl $W_{1}$ is expected to have a total monopole charge of $-8$ after considering SOC. Nevertheless, our calculation shows that the $e_{g}$ orbitals forming the twofold quadruple Weyl node will evolve into a spin-$\frac{3}{2}$ Weyl node with a monopole charge of $+4$. Thus, the monopole charge for the Weyl node $W_{1}$ at $\Gamma$ point changes both the chirality and value after considering SOC, and we will explore this puzzle in the next section. 
Figure~\ref{fig:FIG_1} (c) also shows the approximate positions for Weyl nodes after considering SOC both in the bulk BZ and in the surface BZ along [001] direction. In the spinful case, the black dot at $\Gamma$/$\bar{\Gamma}$ is the spin-$\frac{3}{2}$ Weyl node with a monopole charge of $+4$.
\begin{center}
\begin{figure}
\includegraphics[scale=0.9]{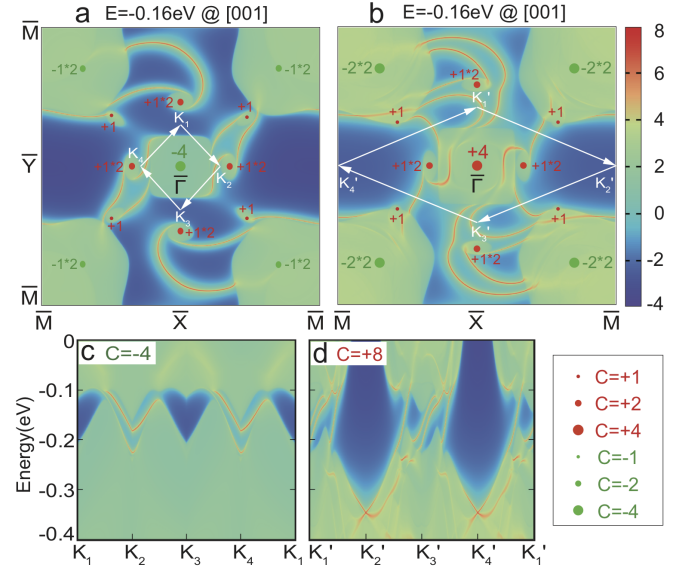}
\caption{ (a-b) Fermi arcs for BaIrP in the absence and in the presence of spin-orbit coupling, and both of them are calculated at -0.16eV along [001] direction. (c-d) are the calculated surface states along a closed loop K$_{1}$-K$_{2}$-K$_{3}$-K$_{4}$-K$_{1}$ and K$_{1}'$-K$_{2}'$-K$_{3}'$-K$_{4}'$-K$_{1}'$, which are marked in white lines in (a-b).
\label{fig:FIG_3}}
\end{figure}
\end{center}

\begin{center}
\begin{figure*}
\includegraphics[scale=1.02]{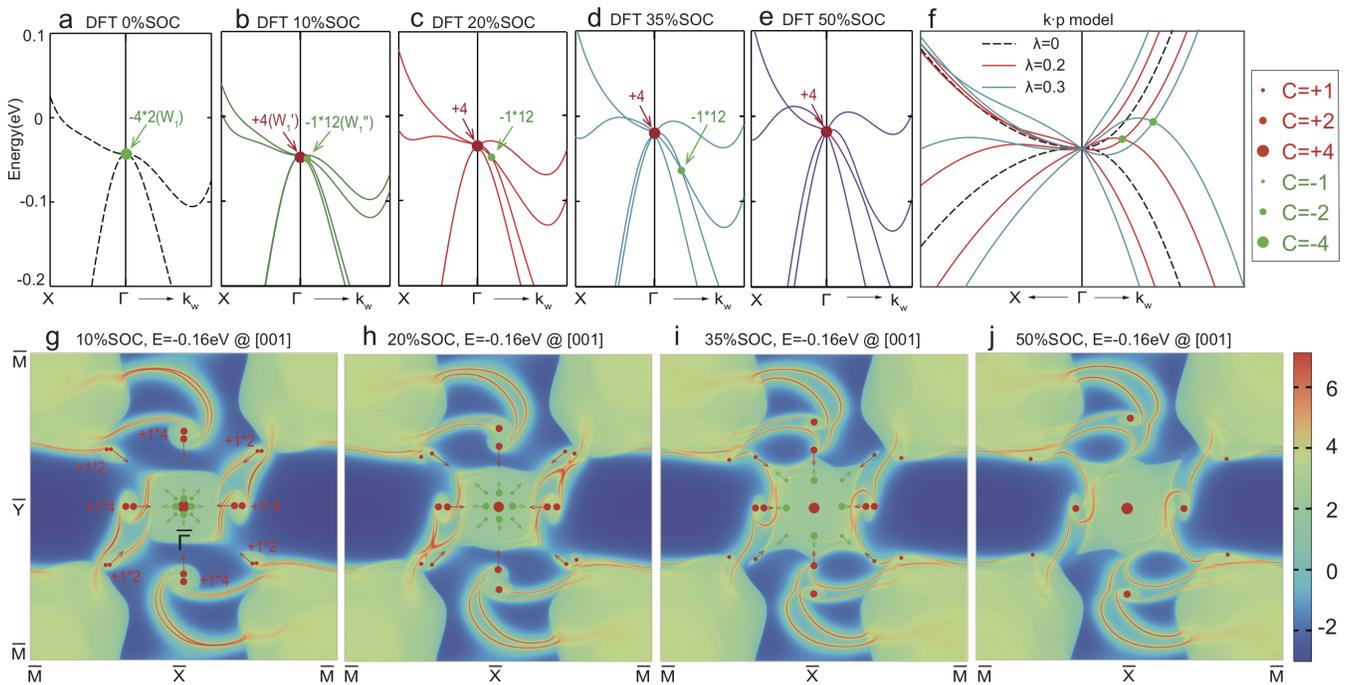}
\caption{ (a-e) are the band structures of BaIrP along X-$\Gamma$-$k_{w}$ direction by DFT calculations with different magnitudes of SOC, where $k_{w}$ is the momentum of Weyl node $W_{1}''$ generated from the quadruple Weyl node at $\Gamma$. (f) is the band structure obtained from the $k\cdot p$ model by considering different magnitudes of SOC, $i.e.$, different values of $\lambda$. (g-j) are calculations of Fermi arcs at 0.16eV along [001] direction, for different values of SOC. Once SOC is turned on, twofold quadruple Weyl node $W_{1}$ at $\Gamma$ point will evolve into a fourfold quadruple Weyl node $W_{1}'$ with 12 Weyl nodes $W_{1}''$ around, and 12 Weyl nodes $W_{3}$ will evolve into 12 Weyl nodes $W_{3}'$ and 12 Weyl nodes $W_{3}''$. As the value of SOC increases, the Weyl nodes $W_{1}''$ and $W_{3}''$ will move close to each other, which are labelled by arrows in (g-i), and are annihilated with each other when SOC is 50\% of the original value.
\label{fig:FIG_4}}
\end{figure*}
\end{center}
\subsection{Surface states for Chiral fermions in BaIrP}
An open Fermi arc is one of the unique features of Weyl semimetals, and the ends of the arcs are pinned to the projection of Weyl nodes from the bulk. To figure out the Fermi arc configuration for the quadruple Weyl fermions, we calculated the surface local density of states~ (LDOS), which is the imaginary part of surface Green's function\cite{CAL_GREEN1,CAL_GREEN2}, along [001] direction. Figure~\ref{fig:FIG_3} (a-b) are the LDOS at -0.16eV along [001] direction without/with SOC.  

On the [001] surface, $\mathcal{T}$ and the in-plane translation symmetry are the only symmetries left. In the spinless case, the quadruple Weyl $W_{1}$ node at $\Gamma$ point will be projected to $\bar{\Gamma}$, and the Weyl node $W_{2}$ along $\Gamma$-R direction will be projected to $\bar{\Gamma}$-$\bar{M}$ direction, as shown in Fig.~\ref{fig:FIG_3} (a). The Weyl nodes $W_{3}$ will be projected to different positions on the surface BZ, but around the quadruple Weyl node at $\bar{\Gamma}$. 
According to the bulk-boundary correspondence, there will be four Fermi arcs coming out from the quadruple Weyl node at $\bar{\Gamma}$, and connected to the Weyl nodes with different chirality, as shown in  Fig.~\ref{fig:FIG_3} (a). After taking SOC into consideration, the twofold quadruple Weyl $W_{1}$ at $\bar{\Gamma}$ will evolve into a fourfold quadruple Weyl node $W_{1}'$, and its monopole charge changes to positive at the same time. As shown in Fig.~\ref{fig:FIG_3} (b), all the Fermi arcs are coming from the fourfold quadruple Weyl node $W_{1}'$ and the Weyl node $W_{3}'$, connecting to the quadratic double Weyl node along $\bar{\Gamma}$-$\bar{M}$ direction. 

In order to show the relationship between the nonzero Chern number and the number of Fermi arcs, we calculate the surface states along a clockwise loop surrounding $\bar{\Gamma}$, both in the spinless case and in the spinful case. Surface states shown in Fig.~\ref{fig:FIG_3} (c) are along the loop K$_{1}$-K$_{2}$-K$_{3}$-K$_{4}$-K$_{1}$, which surrounds only the twofold quadruple Weyl node $W_{1}$, as marked in Fig.~\ref{fig:FIG_3} (a). Thus, there are four surface states connecting from the conduction band to the valence band, which demonstrate the existence of the quadruple Weyl node with a Chern number of $-4$. Surface states shown in Fig.~\ref{fig:FIG_3} (d) are along a larger loop K$_{1}'$-K$_{2}'$-K$_{3}'$-K$_{4}'$-K$_{1}'$, which includes the fourfold quadruple Weyl node $W_{1}'$ and the four Weyl node $W_{3}'$, as shown in Fig.~\ref{fig:FIG_3} (b). Thus, there are eight surface states connecting from the valence band to the conduction band, which demonstrates a C=$+8$ spectral flow.

\subsection{Evolution for Quadruple Weyl fermions}
As we discussed above, both the value of monopole charge and the chirality of the quadruple Weyl node $W_{1}$ have changed after considering SOC. To figure out the influence of SOC in this process, we gradually increase the value of SOC in the DFT calculation. 
Figure~\ref{fig:FIG_4} (a) shows the dispersion of BaIrP along X-$\Gamma$-$k_{w}$ direction with the twofold quadruple Weyl node without SOC, where $k_{w}$ is the position for the Weyl node $W_{1}''$ generated from $\Gamma$ point. 
After turning on the SOC, we find that the twofold quadruple Weyl carrying a monopole charge of $-4$ will become a fourfold one carrying a monopole charge of $+4$, and generate 12 spin-$\frac{1}{2}$ Weyl fermions $W_{1}''$ carrying a monopole charge of $-1$. When the value of SOC becomes larger, the spin-$\frac{1}{2}$ Weyl node $W_{1}''$ will move far from $\Gamma$ point and be annihilated with 12 spin-$\frac{1}{2}$ Weyl node $W_{3}''$ when the value of SOC is 50\% of the original one. 

In order to have a better understanding for this process, we also calculated the Fermi arcs along [001] direction with different strength of SOC, which are shown in Fig.~\ref{fig:FIG_4} (g-j). Figure~\ref{fig:FIG_4} (g) shows the Fermi arcs with 10\% of the original SOC, where the Weyl node $W_{3}$ will be doubled into $W_{3}'$ and $W_{3}''$ with a monopole charge of $+1$ for each node, and the twofold quadruple Weyl node $W_{1}$ with a monopole charge of $-4$ will evolve into a fourfold quadruple Weyl node $W_{1}'$ with a monopole charge of $+4$ and 12 Weyl node $W_{1}''$ around with a monopole charge of $-1$ for each node. 
When the strength of SOC become larger, the Weyl nodes $W_{1}''$ and $W_{3}''$ with different monopole charge will move closer to each other along the directions marked by the arrows, and be annihilated when the value of SOC is 50\% of the original one.

\subsection{$k\cdot p$ analysis for Quadruple Weyl fermions}
\label{IIF}
In the absence of SOC, the low-energy effective $k\cdot p$ model for the spin-up electrons is a two-dimensional Hamiltonian: 
$H_{nsoc} (\mathbf{k})=-\begin{pmatrix}
Ak_{x}k_{y}k_{z} & B (k_{x}^{2}+\omega k_{y}^{2}+\omega^{2}k_{z}^{2}) \\ 
B (k_{x}^{2}+\omega^{2}k_{y}^{2}+\omega k_{z}^{2}) & -Ak_{x}k_{y}k_{z} 
\end{pmatrix}$, where $\omega$=$e^{-\frac{2\pi i}{3}}$, and $A$ and $B$ are real constants. Details for the derived process are in the supplementary. %
{The twofold degenerated quadruple Weyl node has a monopole charge of $-4$, which is the largest monopole charge that a twofold degenerated Weyl can have in crystalline materials. It has a $\mathbf{k}^{3}$ dispersion along the [111] direction and $\mathbf{k}^{2}$ dispersion along other directions. Such kind of high-order terms in the dispersion will lead to a larger density of states than other unconventional Weyl fermions, which always have a linear term in $\mathbf{k}$ and contributes less topological carriers.}

After considering SOC, the $k\cdot p$ model at $\Gamma$ is derived up to the linear order in $\mathbf{k}$ as: 
\begin{widetext}
\begin{eqnarray*} 
H_{soc} (\mathbf{k})=\begin{pmatrix}
ck_{z}                    & ck_{-}                      & (a-ib)\omega^{2}k_{z}     & (a-ib) (k_{x}-i\omega k_{y}) \\
ck_{+}                    & -ck_{z}                    & (a-ib) (k_{x}+i\omega k_{y})           & - (a-ib)\omega^{2}k_{z} \\
 (a+ib)\omega k_{z}& (a+ib) (k_{x}-i\omega^{2}k_{y})  &ck_{z}& ck_{-}\\
 (a+ib) (k_{x}+i\omega^{2}k_{y}) & - (a+ib)\omega k_{z}& ck_{+}& -ck_{z}
\end{pmatrix},
\end{eqnarray*}
\end{widetext}
where $a$, $b$ and $c$ are real constants, with the details given in the supplementary. 
This Hamiltonian $H_{soc}$ represents the SOC itself, because in the absence of SOC, the $k\cdot p$ Hamiltonian up to the linear order in $\mathbf{k}$ is zero. Therefore, in order to study the influence of SOC on the quadruple Weyl fermions, we can mix $H_{nsoc}$ and $H_{soc}$ with a parameter $\lambda$, like $H (\mathbf{k})=H_{nsoc} (\mathbf{k})\otimes \sigma_{0} + \lambda\cdot H_{soc} (\mathbf{k})$. When $\lambda$=0, $H (\mathbf{k})$ corresponds to a double twofold quadruple Weyl node $W_{1}$ with a total monopole charge of $-8$. After increasing the value of $\lambda$, $H (\mathbf{k})$ will have a fourfold quadruple Weyl node $W_{1}'$ with a monopole charge of $+4$ and 12 spin-$\frac{1}{2}$ Weyl nodes $W_{1}''$carrying a monopole charge of $-1$ around. Figure~\ref{fig:FIG_4} (f) shows the band structure of $H (\mathbf{k})$ with $A=B=1$, $a$=$-\frac{\sqrt{3}}{2}$, $b$=$\frac{1}{2}$ and $c$=$\frac{3}{8}$. The 12 spin-$\frac{1}{2}$ Weyl nodes $W_{1}''$ will move far from $\Gamma$ when the value of $\lambda$ increases, which matches the DFT calculations very well.  

\subsection{Discussions on physical properties of quadruple Weyl fermions}
Quadruple Weyl fermions in BaIrP are also Kramers Weyl fermions, which means that a $k\cdot$I$_{N\times N}$ term is forbidden by time-reversal symmetry, where I$_{N\times N}$ represents a $N$-dimensional identical matrix. Thus, the quadruple Weyl fermion in BaIrP can not be a type-II Weyl fermion in the vicinity of $\Gamma$ point. {Since Kramers Weyl nodes are irreducible representations at TRIM, they cannot be created/annihilated by perturbations, unless the order of irreducible representations at TRIM changes. Furthermore, unconventional Kramers Weyl nodes cannot split into conventional Weyl points by perturbations due to the symmetries in the vicinity of TRIM, although it is allowed by charge conservation.} 
From the nature of Kramers Weyl fermion, quadruple Weyl node $W_{1} (W_{1}')$ is robust and not related to the other Weyl nodes in BaIrP, which means that the quadruple Weyl node in BaIrP will have a different energy with other Weyl nodes. Since the CPGE induced current is proportional to the monopole charge of the Weyl node and the applied optical intensity, the energy offset and large Chern number for the Weyl node make BaIrP and other LaIrSi-type materials ideal platforms to obtain quantized CPGE with a Chern number of $\pm4$. 

Here we would like to note that, the chirality for the quadruple Weyl in LaIrSi-type materials will be changed when the chirality of crystal changes, and so is the chirality of the circularly polarized light in CPGE. Thus, the CPGE effect can help to obtain the chiral properties both for crystal structures and for Weyl nodes.

\begin{center}
\begin{figure}
\includegraphics[scale=0.8]{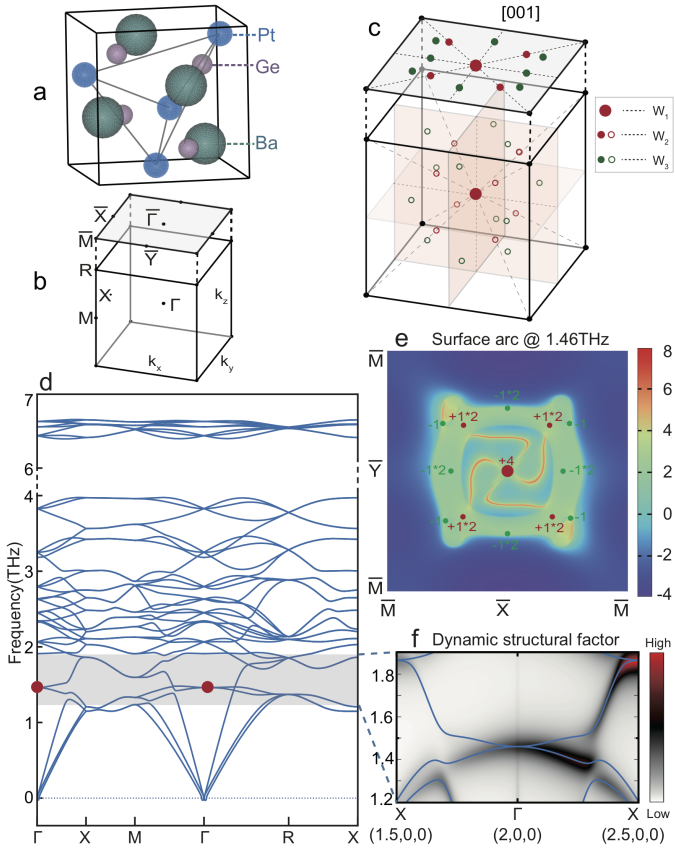}
\caption{ (a) Crystal structure for BaPtGe. (b) Brillouin zone for BaPtGe. (c) Distribution for Weyl nodes in the bulk and surface Brillouin zone. (d) Phonon spectra for BaPtGe. (e) Surface arc for the twofold quadruple Weyl node at 1.46THz. (f) Calculated dynamic structural factor for phonons along X-$\Gamma$-X, crossing the first and second Brillouin zone, where the unit for the momentum is $\pi$/$a$. 
\label{fig:FIG_6}}
\end{figure}
\end{center}

\section{Quadruple Weyl Phonons}
As we discussed above, quadruple Weyl nodes can be obtained in systems preserving both chiral cubic symmetry and $\mathcal{T}$, such as LaIrSi-type materials. 
However, not like in electronic systems where 3 kinds of quadruple Weyl nodes can exist, there is only one kind of quadruple Weyl node in phonon spectra, $i.e.$, twofold quadruple Weyl node. In the following, we will take BaPtGe as an example to discuss the quadruple Weyl nodes. 

\subsection{Quadruple Weyl nodes in BaPtGe}
BaPtGe also belongs to LaIrSi-type materials, with space group of \#198, and the crystal structure is shown in Fig.~\ref{fig:FIG_6} (a). Figure~\ref{fig:FIG_6} (d) shows the phonon spectra for BaPtGe, where the 4th band and 5th band form a quadruple Weyl node at $\Gamma$ point, with Chern number of $C=+4$. 
Like the spinless case in BaIrP, there are also 8 single Weyl nodes $W_{2}$ with Chern number of $C=+1$ along the diagonal high-symmetry $\Gamma-R$ lines, and 12 single Weyl nodes $W_{3}$ with Chern number of $C=-1$ located on the $k_{x,y,z}$=0 planes, as shown in Fig.~\ref{fig:FIG_6} (c). Positions for those three kinds of Weyl nodes are listed in Tab.~\ref{tab:Tab2}.

To figure out the surface arcs connection between the quadruple Weyl node and single Weyl nodes, we calculated the LDOS at 1.46THz along [001] direction, as shown in Fig.~\ref{fig:FIG_6} (e). Due to the surface-bulk correspondence, there are 4 surface arcs coming from the quadruple Weyl node $W_{1}$ at $\Gamma$ point, and connecting to the Weyl nodes $W_{3}$. Because $\mathcal{T}$ and the in-plane translation symmetry are the only symmetries left on the [001] surface, the surface arcs are related by $\mathcal{T}$ symmetry.


\subsection{Dynamical structural factor}
In addition to the nontrivial surface arcs, directly probing on the bulk bands can also help to demonstrate the existence of quadruple Weyl nodes in the phonon spectra. For example, in the measurements of neutron scattering and inelastic X-ray scattering (IXS), the intensity of phonon dynamic structure factor (DSF) highly depends on the vibrational mode and momentum transfer. Thus, intensity for the bands composing the quadruple Weyl node will be different at different momenta. Moreover, the intensity will also be different at the same momenta in the different BZ. 

Figure~\ref{fig:FIG_6} (f) is the DSF calculation simulating IXS measurements along X-$\Gamma$-X direction, which extends over two BZ. Although the phonon dispersions are the same along X (1.5,0,0)-$\Gamma$ (2,0,0) and $\Gamma$ (2,0,0)-X (2.5,0,0) direction, which has a large splitting between the two bands forming the quadruple Weyl node, the DSF is highly different due to the different momentum transfer. Thus, by detecting the DSF at different momenta, we can directly trace out the topological nature of quadruple Weyl node in BaPtGe.

\begin{center}
\begin{figure}
\includegraphics[scale=0.8]{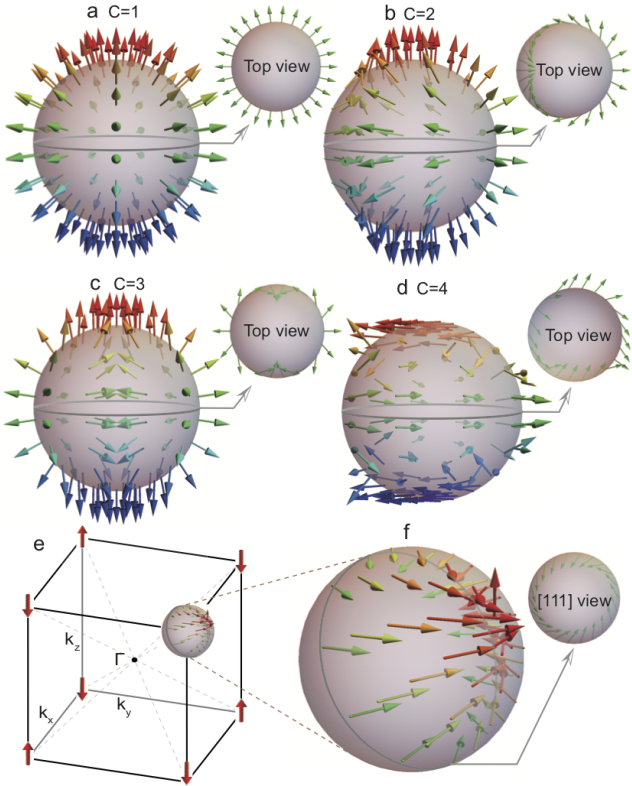}
\caption{ (a-c) Pseudospin structure for twofold Weyl nodes with Chern number of $+1,+2,+3$ on a sphere. Here, we use the Hamiltonian in Sec.\ref{IIB} with $N=1,2,3$ as an example. (d) Pseudospin structure for a twofold quadruple Weyl node with Chern number of $+4$ on a sphere. Here, we use the Hamiltonian H$_{nsoc}$ in Sec.\ref{IIF}. (e) $\hat{S_{z}}$ component for twofold quadruple Weyl node at eight momenta along diagonal lines. (f) Pseudospin on a half sphere along diagonal direction, which contributes a half wrapping number. Four insets in the upper right of (a-d) are the top view of pseudospin structures on the circle located at $k_{z}=0$ plane, and the inset in (f) is the side view along [111] direction. The pseudospin wraps the unit sphere differently for systems with different Chern numbers.
\label{fig:FIG_7}}
\end{figure}
\end{center}
%
\begin{table*}[]
\begin{tabular}{cccc}
\hline 
Positions (2$\pi$/$a$) & $W_{1}$  & $W_{2}$   & $W_{3}$   \\
\hline \hline %
Monopole charge     & $+4$  & $+1$    &  $-1$ \\
\hline 
Positions   & (0,0,0)      & \begin{tabular}[c]{@{}c@{}}$\pm$ (0.1824,0.1824,0.1824) $\pm$ (-0.1824,-0.1824,0.1824) \\ $\pm$ (-0.1824,0.1824,-0.1824) $\pm$ (0.1824,-0.1824,-0.1824) \end{tabular} & \begin{tabular}[c]{@{}c@{}}$\pm$ (0.21,0.235,0) $\pm$ (0,0.21,0.235) $\pm$ (0.235,0,0.21) \\  $\pm$ (0,-0.21,0.235) $\pm$ (0.21,-0.235,0)  $\pm$ (-0.235,0,0.21) \end{tabular} \\ \hline 
\end{tabular}
\caption{Positions for the Weyl nodes $W_{1}$, $W_{2}$, and $W_{3}$ in BaPtGe, where $a$ is the lattice constant.}
\label{tab:Tab2}
\end{table*}

{\subsection{Pseudospin for quadruple Weyl nodes}}
As a spin-0 particle, phonons do not possess any spin properties. However, we can study the relationship between the pseudospin texture and the Chern number for Weyl nodes by using the low-energy effective $k\cdot p$ model $H_{2\times2} (\mathbf{k})=f_{i=x,y,z} (\mathbf{k})\cdot\sigma_{i=x,y,z}$, where $\sigma_{i}$ is the Pauli matrix and the parameters $f (\mathbf{k})$ determine not only the Chern number, but also the pseudospin texture. 
Here, we define a pseudospin to be $\hat{S_{i}}\equiv\frac{f_{i} (\mathbf{k})}{|f_{i} (\mathbf{k})|}$, and it is in fact an expectation value of $\sigma_{i}$ for the upper band: $\hat{S_{i}}=<\psi_{upper}|\sigma_{i}|\psi_{upper}>$, where $\psi_{upper}$ denotes the wavefunction for the upper band forming the Weyl node. 
For example, Figure~\ref{fig:FIG_7} (a) shows the pseudospin configuration for a spin-$\frac{1}{2}$ Weyl node. 
When there is a gap everywhere on the sphere enclosing the Weyl node, the way that the pseudospin wrapping the sphere defines a wrapping number $M$. 

The wrapping number is defined by an integral $M=\frac{1}{8\pi}\int_{S}\hat{\mathbf{S}}\cdot (\frac{\partial \hat{\mathbf{S}}}{\partial k_{i}}\times\frac{\partial \hat{\mathbf{S}}}{\partial k_{j}})\cdot \epsilon_{ijk} \hat{n}_{\mathbf{k}}dS_{\mathbf{k}}$ over a closed surface enclosing the Weyl node, and $dS_{\mathbf{k}}$ denotes the surface element. 
This is an integer topological index characterizing a mapping from the two-dimensional closed surface $S$ to the pseudospin $\hat{\mathbf{S}}$ taking a value on a unit sphere. 
Thus, this wrapping number $M$ represents how many times the pseudospin sweeps out the sphere. 
In fact, this wrapping number for the pseudospin is equal to the Chern number, $i.e.$, $M=C$. 
Therefore, for Weyl nodes with Chern number $C=\pm1$, pseudospin $\hat{\mathbf{S}}$ will take every value once on the sphere, as shown in Fig.~\ref{fig:FIG_7} (a). For other twofold degenerated Kramers Weyl nodes with Chern number of $\pm2$, $\pm3$, and $\pm4$, the pseudospin will cover the Weyl node twice, three and four times, as shown in Fig.~\ref{fig:FIG_7} (b-d). 
 In order to show the wrapping process for the pseudospin texture more clearly, we calculated the pseudospin texture on a circle located at $k_{z}=0$ plane for each system. As shown in the upper-right insets of Fig.~\ref{fig:FIG_7} (a-d), pseudospin texture on the circle wraps differently in systems carrying a different Chern number. 
 
Since pseudospin around the twofold quadruple Weyl node proposed in our paper wraps the unit sphere four times in a complex way, as shown in Fig.~\ref{fig:FIG_7} (d), we will offer an intuitive perspective to understand it. According to the  Hamiltonian H$_{nsoc}$ in Sec.\ref{IIF}, pseudospin components are $\hat{S_{x}}=(k_{z}^{2}-\frac{1}{2}(k_{x}^{2}+k_{y}^{2}))/|f_{x}(\mathbf{k})|$, $\hat{S_{y}}=\frac{\sqrt{3}}{2}(k_{x}^{2}-k_{y}^{2})/|f_{y}(\mathbf{k})|$ and $\hat{S_{z}}=k_{x}k_{y}k_{z}/|f_{z}(\mathbf{k})|$. Due to the existence of $C_{3}$ rotation in chiral cubic systems, $\hat{S_{x}}$ and $\hat{S_{y}}$ will be zero along four diagonal directions, leaving the pseudospin component $\hat{S_{z}}$ pointing at out-of-plane $\pm k_{z}$ direction, as shown in Fig.~\ref{fig:FIG_7} (e). However, when $\mathbf{k}$ is small enough, and away from four diagonal lines, $\hat{S_{x}}$ and $\hat{S_{y}}$ components will contributes more than $\hat{S_{z}}$ component, and makes the pseudospin lies in the $k_{x}k_{y}$-plane. 
Thus, contribution to the wrapping number from the vicinity of the eight points along diagonal directions will depend on two parameters according to the definition formula, $i.e.$, the sign of $\hat{S_{z}}$ component at the eight points and the wrapping phase $\theta\ (=\pm 2\pi)$ of pseudospin ($\hat{S_{x}}$, $\hat{S_{y}}$) nearby. Those two parameters are similar with $vorticity$ and $helicity$ of skyrmions, which are needed to define the skyrmion wrapping number\cite{nii2015uniaxial,garaud2015properties,koshibae2014creation}. 
The sign of $k_{x}k_{y}k_{z}$ determines the sign of pseudospin component $\hat{S_{z}}$ at $\mathbf{k}$ being in the diagonal directions, as labeled in Fig.~\ref{fig:FIG_7} (e), which has a positive sign at the $(k,k,k)$ momentum with $k>0$. Pseudospin on the circle perpendicular to the [111] direction are shown in the inset of Fig.~\ref{fig:FIG_7} (f), which has a ${2}{\pi}$ wrapping phase. Thus, the pseudospin around the $(k,k,k)$ momentum contributes to a $\frac{1}{2}$ wrapping number. Namely, the contribution from the vicinity of a point in the diagonal direction is given by {$\hat{S_{z}}\cdot \frac{\theta}{2\pi}\cdot \frac{1}{2}$ (=$\pm\frac{1}{2}$). }
In the vicinity of the $(-k,-k,-k)$ momentum, although the sign of pseudospin component $\hat{S_{z}}$ is negative, pseudospin on the circle perpendicular to the [$\bar{1}\bar{1}\bar{1}$] direction will have a $-{2}{\pi}$ wrapping phase, which will also contribute a $\frac{1}{2}$ wrapping number. 
Due to the existence of $C_{2x,2y,2z}$ rotation symmetry in chiral cubic crystals, other six momenta along diagonal directions will also contribute a $\frac{1}{2}$ wrapping number. 
Thus, the wrapping number in this system will be $+4$ in total.

\section{Methods}
All the band structures are implemented in Vienna $ab$ $initio$ package (VASP)\cite{DFT_VASP1,DFT_VASP2,DFT_VASP3,DFT_VASP4} within the Perdew-Berke-Ernzerhof (PBE) exchange correlation\cite{CAL_VASP}. For the electronic structure calculations, a k-point mesh of $7\times7\times7$ is used in the self-consistent calculation, and the cut-off energy for the plane-wave basis is 450 eV. WANNIER90 package\cite{wannier90} is used for calculating the Wannier function based numerical tight-binding Hamiltonian, which is used for surface states and Fermi arcs calculations. 
For the phonon spectra calculation, density functional perturbation theory is used after a crystal structure relaxation with a condition that residual force on each atom is less than 0.01 eV$\AA^{-1}$. 
Chern numbers are calculated by the Wilson-loop method\cite{CAL_WILSON1,CAL_WILSON2}, and the crystal structures are obtained from inorganic crystal structure database (ICSD)\cite{belsky2002new}.

\section{Conclusion}
We proposed a new kind of unconventional Kramers Weyl node, $i.e.$, twofold degenerated quadruple Weyl node with Chern number of $\pm4$ for each band that has been missed in the previous studies. A series of LaIrSi-type materials are proposed in our paper that have such kind of twofold quadruple Weyl node in the spinless electronic band structure and phonon spectra. After considering SOC, the twofold quadruple Weyl node will evolve into a fourfold quadruple Weyl node with several spin-$\frac{1}{2}$ Weyl nodes around. As a member of Kramers Weyl fermions, the twofold degenerated quadruple Weyl node also preserves exotic transport and optical phenomena, among which we would like to note that LaIrSi-type materials are ideal candidates for quantized circular photogalvanic effect. This is because in LaIrSi-type materials, the Kramers Weyl node has a large monopole charge and there is a big energy offset between the Weyl nodes with different chirality\cite{QCPE_NC,asteria2019measuring}. In the discussion of quadruple Weyl phonon, we also discuss the relationship between the monopole charge and pseudospin wrapping number for Weyl nodes with different Chern numbers. Among them, the twofold quadruple Weyl node has an exotic pseudospin texture, and we offered a new and intuitive way to understand it. 

\section{Acknowledgements}
T.T. Z. and S. M. acknowledge the supports from Tokodai Institute for Element Strategy
 (TIES) funded by MEXT Elements Strategy Initiative to
Form Core Research Center. S. M. also acknowledges support by JSPS KAKENHI Grant Number JP18H03678. 
C. F. acknowledges support from the Ministry of Science and Technology of China under grant numbers 2016YFA0302400 and 2016YFA0300600; the National Science Foundation of China under grant numbers 11674370; and the Chinese Academy of Sciences under grant numbers XXH13506-202 and XDB07020100.

\newpage
\begin{center}
\begin{figure*}
\includegraphics[scale=0.8]{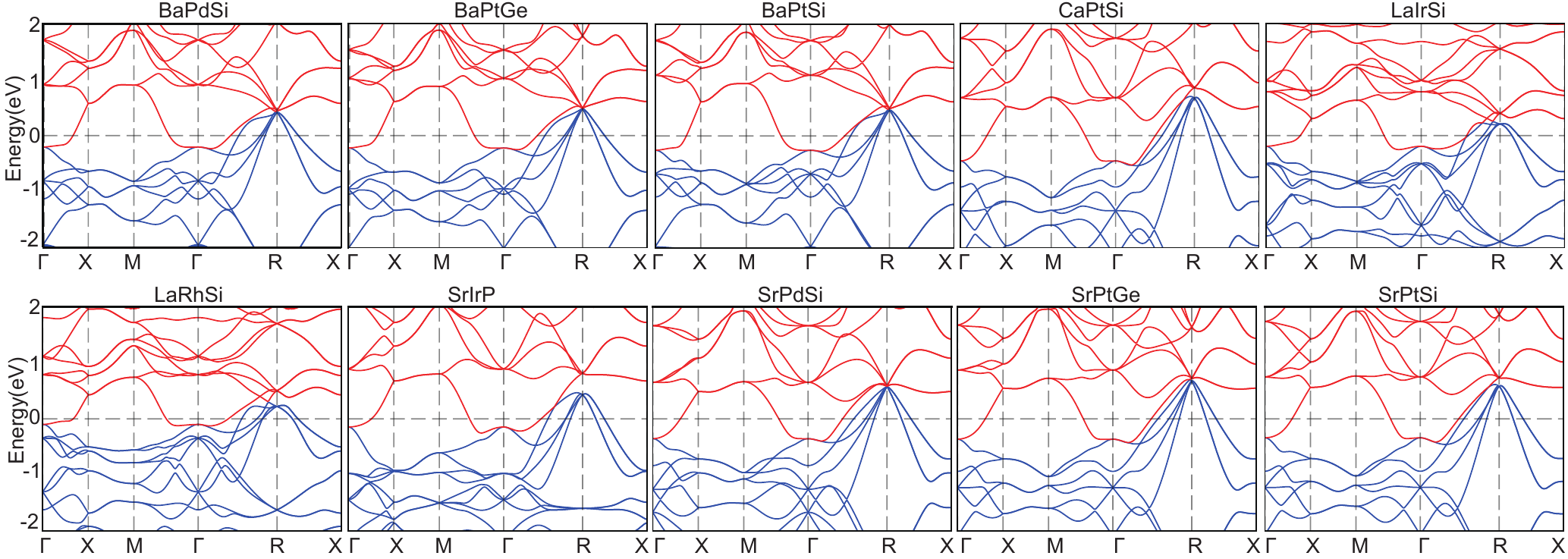}
\caption{Band structures for the other 10 materials in LaIrSi-type family in the absence of spin-orbit coupling.
\label{fig:SI_1}}
\end{figure*}
\end{center}

\begin{center}
\begin{figure*}
\includegraphics[scale=0.8]{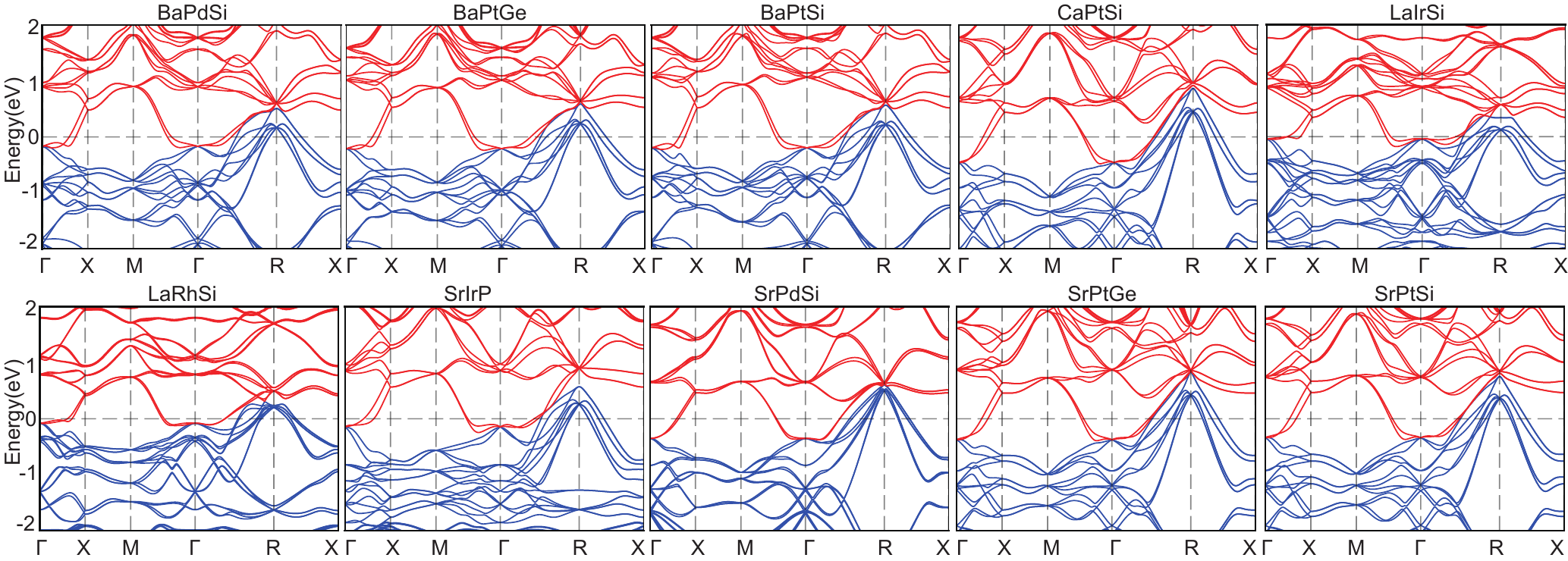}
\caption{Band structures for the other 10 materials in LaIrSi-type family in the presence of spin-orbit coupling.
\label{fig:SI_2}}
\end{figure*}
\end{center}

\section*{SUPPLEMENTARY: Electronic bands for other LaIrSi-type materials}
Figure~\ref{fig:SI_1} and ~\ref{fig:SI_2} show the band structure of other 10 materials for LaIrSi family. They all have a 2-band quadruple Weyl at $\Gamma$ point when the spin-orbit coupling(SOC) is ignored, and 4-band quadruple Weyl when SOC is taken into consideration. All the materials in LaIrSi family are filling-enforced semimetals, which means that there will alway be a quadruple Weyl near the Fermi level regardless of the calculation methods.

\section*{SUPPLEMENTARY: $k\cdot p$ model}
To explore the topological nature of quadruple Weyl Fermions and the change of chirality after considering SOC, we build the low-energy effective $k\cdot p$ mode for further analysis. Generators for \#198 are:

$C_{2x}=\begin{pmatrix}
1 & 0&0 &\frac{1}{2}\\ 0 & -1 &0&\frac{1}{2}\\ 0& 0&-1&0
\end{pmatrix}$, 

$C_{2z}=\begin{pmatrix}
1 & 0&0 &\frac{1}{2}\\ 0 & -1 &0&0\\ 0& 0&-1&\frac{1}{2}
\end{pmatrix}$, 

and $C^{3-}_{\bar{1}1\bar{1}}=\begin{pmatrix}
0 & -1&0 &\frac{1}{2}\\ 0 & 0 &-1&0\\ 1& 0&0&\frac{1}{2}
\end{pmatrix}$.

In the absence of SOC, the $k\cdot p$ model is a two-dimensional one. The irreducible representation matrix for time-reversal symmetry is $\mathcal{T}$=$\sigma_{x}\mathcal{K}$, and that for all the generators are:

$C_{2x}=\begin{pmatrix}
1 & 0 \\ 0 & 1 
\end{pmatrix}$, 

$C_{2z}=\begin{pmatrix}
1 & 0 \\ 
0 & 1 
\end{pmatrix}$,

$C^{3-}_{\bar{1}1\bar{1}}=\begin{pmatrix}
e^{-\frac{i2\pi}{3}} & 0 \\
0 & e^{\frac{i2\pi}{3}}  
\end{pmatrix}$

So low-energy effective model at $\Gamma$ point will be:
$H_{nsoc}(\mathbf{k})=-\begin{pmatrix}
Ak_{x}k_{y}k_{z} & B(k_{x}^{2}+\omega k_{y}^{2}+\omega^{2}k_{z}^{2}) \\ 
B(k_{x}^{2}+\omega^{2}k_{y}^{2}+\omega k_{z}^{2}) & -Ak_{x}k_{y}k_{z} 
\end{pmatrix}$, where $\omega$=$e^{-\frac{2\pi i}{3}}$, A and B are real numbers.

After considering SOC, the irreducible representation matrix for time-reversal symmetry will be $\mathcal{T}=-\sigma_{x}\otimes\sigma_{y}\mathcal{K}$, and that for the generators are:

$C_{2x}=\begin{pmatrix}
0 & -i &0 &0 \\ -i & 0 &0 &0\\ 0 & 0 & 0 & -i \\ 0&0&-i&0
\end{pmatrix}$,

$C_{2z}=\begin{pmatrix}
-i & 0& 0& 0 \\
0 & i & 0& 0 \\
0& 0&-i& 0\\
0& 0& 0& i
\end{pmatrix}$,

$C^{3-}_{\bar{1}1\bar{1}}=\begin{pmatrix}
e^{-i\frac{11\pi}{12}} & e^{-i\frac{11\pi}{12}} & 0& 0 \\
e^{i\frac{7\pi}{12}}  & e^{-i\frac{5\pi}{12}} & 0& 0 \\
0& 0&e^{i\frac{5\pi}{12}} & e^{i\frac{5\pi}{12}} \\
0& 0& e^{-i\frac{1\pi}{12}} & e^{i\frac{11\pi}{12}} 
\end{pmatrix}/\sqrt{2}$,

So the $k\cdot p$ model at $\Gamma$ point is: 
\begin{widetext}
\begin{eqnarray*} 
H_{soc}(\mathbf{k})=\begin{pmatrix}
ck_{z}                    & ck_{-}                      & (a-ib)\omega^{2}k_{z}     & (a-ib)(k_{x}-i\omega k_{y}) \\
ck_{+}                    & -ck_{z}                    & (a-ib)(k_{x}+i\omega k_{y})           & -(a-ib)\omega^{2}k_{z} \\
(a+ib)\omega k_{z}& (a+ib)(k_{x}-i\omega^{2}k_{y})  &ck_{z}& ck_{-}\\
(a+ib)(k_{x}+i\omega^{2}k_{y}) & -(a+ib)\omega k_{z}& ck_{+}& -ck_{z}
\end{pmatrix}
\end{eqnarray*} 
\end{widetext}
which corresponds to a spin-$\frac{3}{2}$ Weyl node with a monopole charge of $+4$. Parameters $a$, $b$ and $c$ are real numbers.

\bibliographystyle{unsrt}
\bibliography{reference}

\end{document}